\documentclass[11pt]{article}
\usepackage{fleqn,cospar}

\usepackage{graphicx}

\def \degmark{$^\circ$}  
\def \nh {$N{\rm _H}$}
\def \src {GRO\thinspace J1655-40} 
\def \Msun {$M_{\odot }$}
\def \rs {$r_{\rm S}$}

\title{SPECTRAL EVOLUTION OF THE CONTINUUM AND 
       DISC LINE IN DIPPING IN \src}

\author{M. Ba\l uci\'nska-Church
\address{School of Physics and Astronomy, 
University of Birmingham, Edgbaston, Birmingham B15 2TT, UK}
\address{Institute of Astronomy, Jagiellonian University, ul. Orla
171, 30-244 Krak\'ow, Poland}
}

\begin{document}
\maketitle

\begin{abstract}
The discovery is reported of emission features in the X-ray spectrum of \src\ 
obtained using {\it Rossi-XTE} on 1997, Feb 26. The features have been fitted
firstly by two Gaussian lines, which in four spectra have average energies of 
$5.85 \pm 0.08$ keV and $7.32 \pm 0.13$ keV, strongly suggestive that 
these are the red and blue shifted wings of an iron disc line from material with velocity 
$\sim $0.33 c. The blue wing is apparently less bright than expected for a disc line 
subject to Doppler boosting, however, known absorption in the spectrum of \src\ at 
energies between $\sim $7 and 8 keV can reduce the apparent brightness of the blue 
wing. The spectra have also been fitted well using the full relativistic disc line 
model of Laor, plus an absorption line. This gives a restframe energy between 6.4 and 
6.8 keV indicating that the line is from highly ionized iron $\rm {K_{\alpha}}$. The 
Laor model also shows that the line originates at radii extending from $\sim $10
Schwarzschild radii (\rs ) outwards. The line is direct evidence for the black hole nature 
of the compact object. The continuum is well described by dominant 
disc blackbody emission plus Comptonized emission. During dipping, spectral evolution 
is well modelled by allowing progressive covering of the disc blackbody and simple
absorption of the Comptonized emission showing that the thermal emission is more 
extended. Acceptable fits are only obtained by including the disc line in the covering 
term, indicating that it originates in the same inner disc region as the 
thermal continuum. Dip ingress times and durations are used to provide 
the radius of the disc blackbody emitter as 170--370 \rs, and the radius of the absorber.
\end{abstract}

\section*{INTRODUCTION}

\src\  (X-ray transient Nova Sco 1994) is a Galactic Jet Source first
discovered in hard X-rays with BATSE (Zhang et al. 1994).  It underwent several 
X-ray outbursts between July 1994 until September 1997 when the source 
switched off in X-rays. High resolution radio observations revealed 
apparently superluminal, relativistic  jets moving in opposite 
directions with velocity of 0.92 c  (Hjellming \& Rupen 1995, Tingay et al. 
1995). However, radio activity of the source is apparently not correlated 
with X-ray outbursts. The source was active at radio wavelengths only for 
approximately 6 months during the early stage of X-ray activity in 1994
(Hjellming \& Rupen 1995).  Shahbaz et al. (1999) used optical 
observations to obtain a
well-constrained mass range for the compact object of 5.5--7.9 \Msun,
leaving no doubt that this is  a black hole.  The inclination of the system 
is in the range of 63.7--70.7\degmark (van der Hooft et al. 1998).  
The black hole is  surrounded by a massive accretion disc filling 
approximately  85\% of the Roche lobe radius (Orosz \& Bailyn 1997).

The source shows strong X-ray dipping, i.e. reductions of X-ray intensity 
caused by obscuration of the X-ray source by absorbing material in the 
line of sight (Fig. 1, left).  The dipping provides a diagnostic of the X-ray emission 
regions and  strongly constrains spectral models because of the requirement that
emission models fit several dip spectra as well as non-dip.
It occurs in this source predominantly in orbital phase 0.68--0.92  
(Kuulkers et al. 2000) suggesting that blobs
of absorbing material are located most probably on the outer rim of the 
accretion disc as in LMXB. 

Recently, a highly red- and blue-shifted iron disc line has been discovered 
in the X-ray spectrum of \src\  (Ba\l uci\'nska-Church \& Church 2000)
from the observation with {\it Rossi-XTE} on 26th February, 1997.  
The observation took place at the beginning of the last outburst seen before the complete
switch-off in X-rays.  The source has not been detected in radio since 1996
despite regular monitoring (Tingay, priv. comm.).
During this observation, strong X-ray dipping occured. In the 
present paper I will discuss the evolution of the continuum and the iron disc 
line during dipping using this data, and will also show how the X-ray dipping 
constrains emission region sizes.

\section*{THE PERSISTENT EMISSION SPECTRUM: DETECTION OF THE DISC LINE}

The observation of \src\ discussed in this paper was made on 1997,
February 26, lasting 14,600~s with an on-source good exposure time of 7,600~s.
Data from the Proportional Counter Array (PCA) instruments are presented.
Four spectra during persistent emission were carefully selected from
regions of the lightcurve where the count rate was particularly stable.
The spectra were fitted with a model consisting of a disc blackbody and 
a power law.  The same model was also used by Zhang et al. (1997).
The disc blackbody
having temperature $kT \sim $1.1 keV was the dominant component contributing 
90\% to the total luminosity  (0.1--100 keV) of 
$\rm {9.6 \times 10^{37}}$ erg $\rm {s^{-1}}$ which is $\sim $10\% of the Eddington luminosity
for a 7\Msun\ object.
The power law was very steep with the photon index $\Gamma \sim2.4$.
The continuum parameters agree well with the results published by 
Kuulkers et al. (1998).
Although the above model gave the best fit to the continuum, the fits
were generally poor ($\chi^2/{\rm dof} \sim$133/91) with systematic 
residuals at $\sim$5.8 keV and $\sim$7.3 keV in every spectrum analysed 
(Fig. 1, right).

\begin{figure}
\includegraphics[width=90mm, height=87mm]{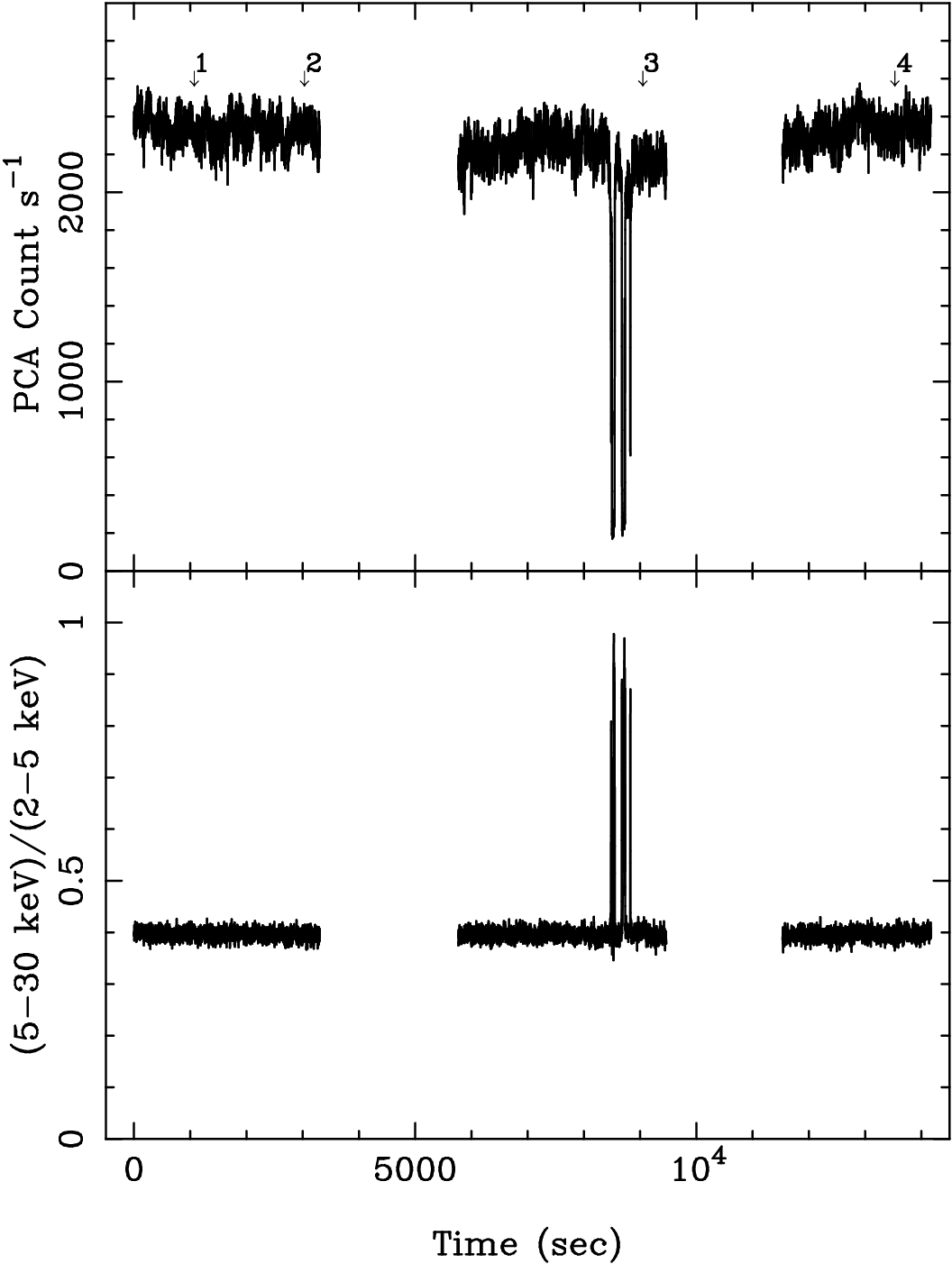}
\includegraphics[width=90mm, height=87mm]{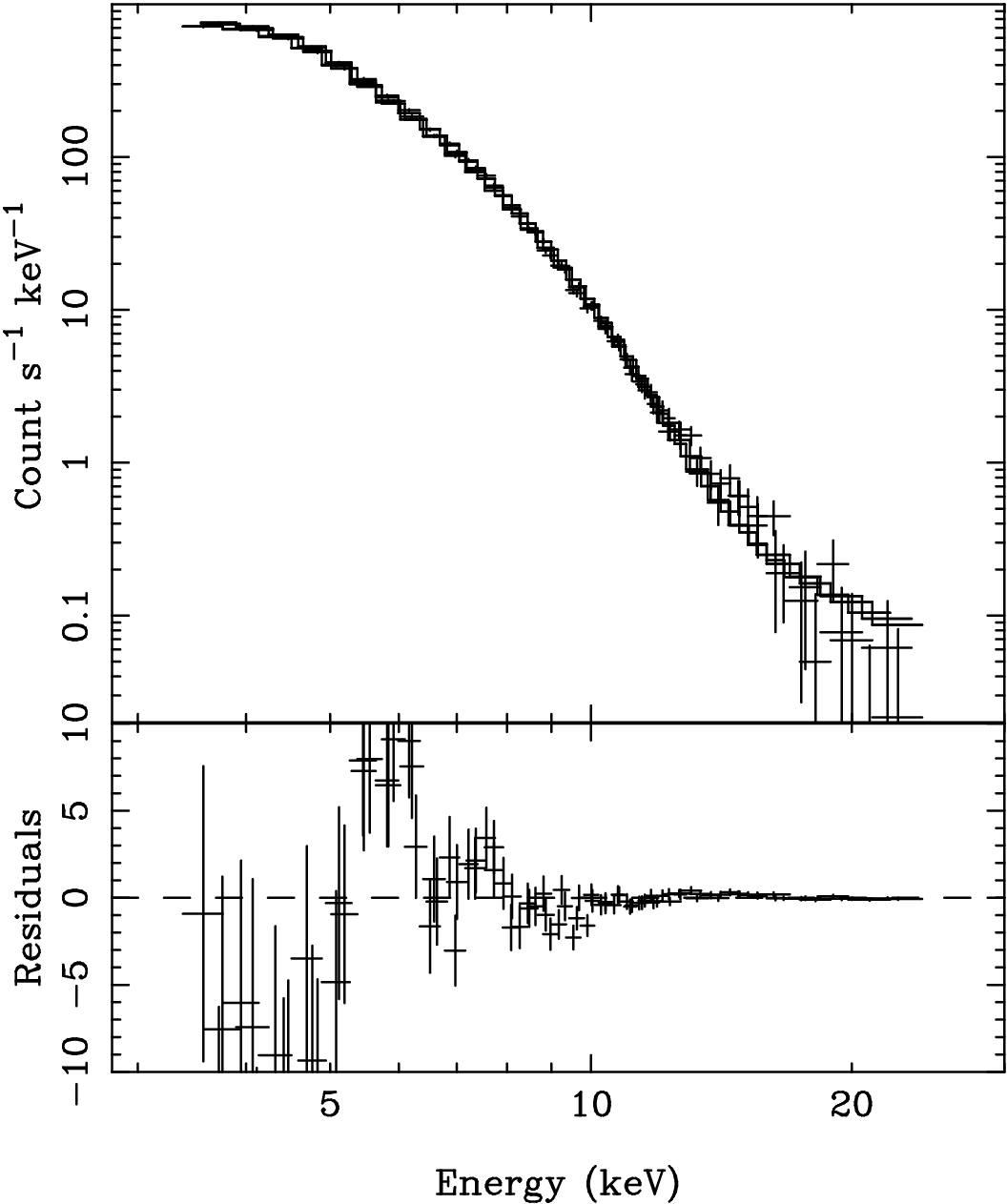}
\vskip -9 mm
\caption{Left: PCA light curve in the energy band 2--30 keV 
with 16-s binning 
after deadtime correction. Right: Spectrum of a single 
data set with residuals; 
data from 3 PCUs are shown, and were fitted simultaneously.}
\end{figure}%

By adding two Gaussian lines to the continuum, the fits became very
good with $\chi^2/{\rm dof} \sim$70/85, and an F-test showed that 
the fits were improved with significance 
$>>$ 99.9\%  (Fig. 2, left).  
The mean line energies of the four spectra are $\bar{E}_1$ = 
$5.85 \pm 0.08$ and $\bar{E}_2 = 7.32 \pm 0.13$ keV.  If we 
assume that the splitting of the two lines is caused
by the relativistic Doppler shift, the rest energy of the line is 
$E_{\rm rest}$ = $6.88 \pm 0.12$ keV and the velocity $v/c = \beta $ is 
0.33 for an inclination
$\hat {\imath}$ = 70\degmark .  However, the intensity of the ``blue''
line is always  smaller than the intensity of the ``red'' line.
Therefore, the intensities of the two lines observed cannot simply be 
due to a disc line where Doppler boosting of the blue line is expected.
However, in several {\it ASCA} spectra of \src\, iron absorption-line
features were found at energies which will influence the observed
strength of the blue line (Ueda et al. 1998).
 
\begin{figure}
\includegraphics[width=90mm, height=87mm]{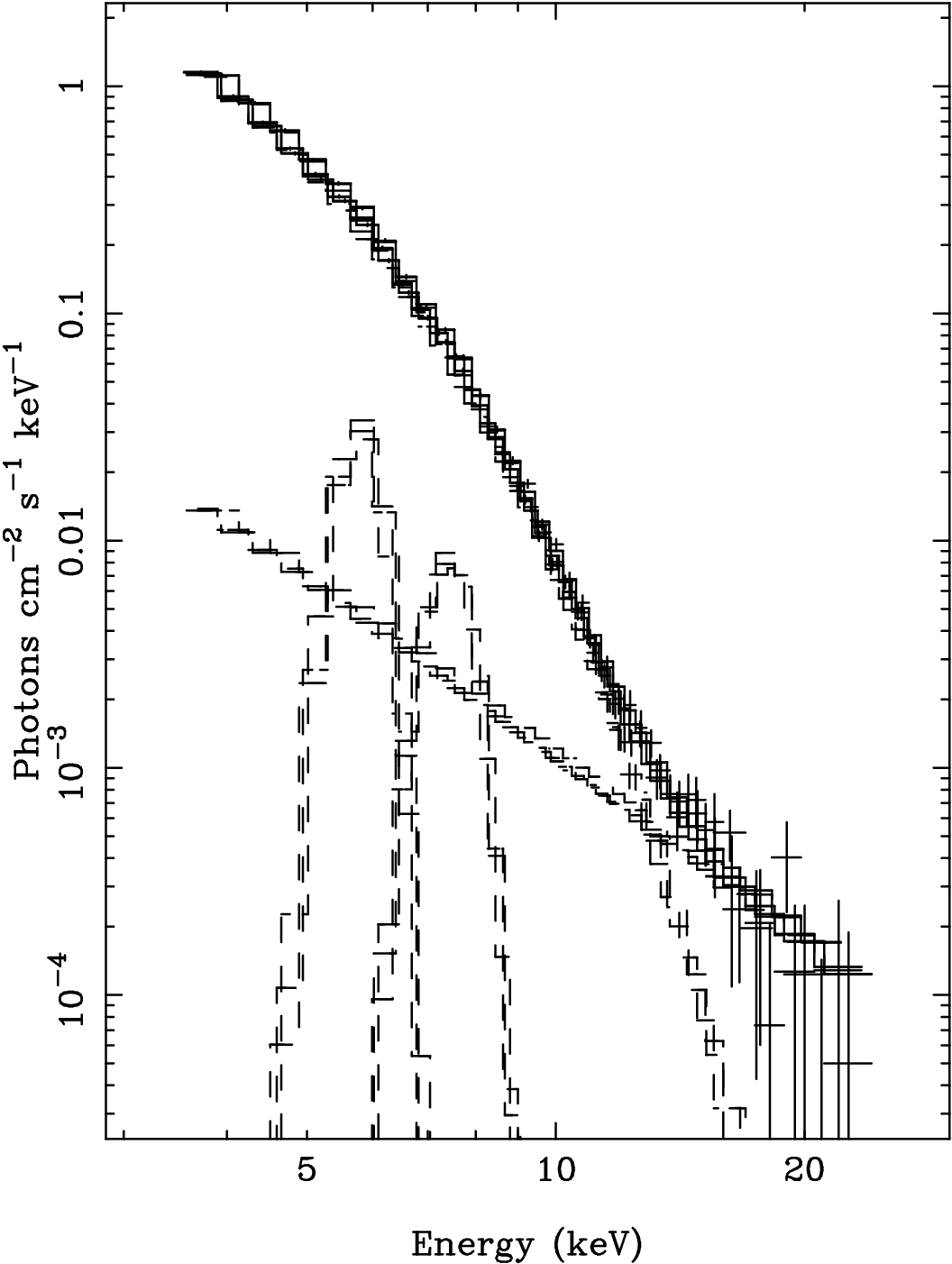}
\includegraphics[width=90mm, height=87mm]{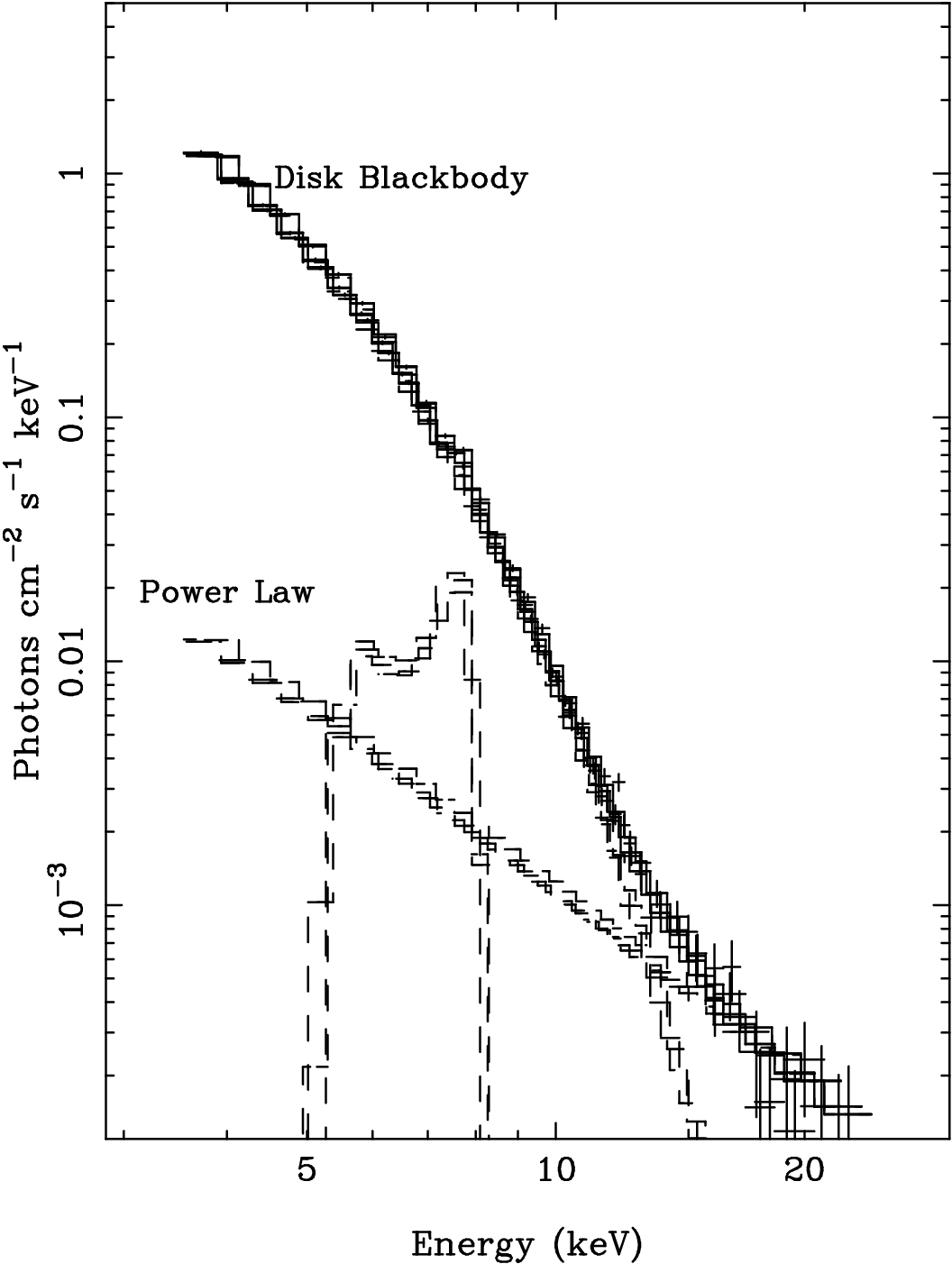}
\vskip -9 mm
\caption{Left: Unfolded spectrum of the best fit to a single spectrum, 
including two Gaussian lines. Right: Best-fit with a Laor-disc line + absorption 
line model. Two continuum components and the Laor line shown; the absorption line
cannot be plotted as it has negative normalization.}
\end{figure}%

Consequently,  we carried out spectral fitting with the continuum described
as before and a Laor disc line + an absorption line added.  Good quality fits 
($\chi^2/{\rm dof} \sim76/84$) were
obtained for all spectra with results as follows.
The mean value of rest-frame energy of the emission line was $6.56 \pm 0.14$ 
keV corresponding to highly ionized Fe (XXII) K$_\alpha $ emission.
However, the energy varied from 6.4 keV to 6.8 keV between spectra
not constraining very well the ionization state.  The inner radius of the emission region 
$r_1$ was found to be $\sim $10 \rs\ where \rs\ is 
the Schwarzschild radius ($19.9 \pm 3.6$ km for the possible range of black 
hole mass in this source); the outer radius $r_2$ was poorly constrained 
to $\geq$ 50 {\rs}. The inner radius value being larger than the radius 
of the last stable orbit suggests that the inner part of the disc up to 10 \rs\
is totally ionized and does not contribute to the emission line.  The mean energy 
of the absorption line from the 4 spectra is well constrained at $7.09 \pm 0.13$ keV 
corresponding to the Fe K$_\alpha $ or  Fe K$_\beta $ line and occurs at
the same energy as the blue wing of the disc line.  Thus it is
quite possible to fit the observed features by a combination of a Laor
disc line with absorption (known to take place in the source), and the
emission line has the expected profile with a boosted blue wing (Fig. 2, right).
Iron disc lines have since been detected in 4U\thinspace 1630-47 (Cui et al. 2000) and 
XTE\thinspace J1748-288 (Miller et al. 2000).

\section*{SPECTRAL EVOLUTION IN DIPPING: THE LINE AND CONTINUUM}
 
During the {\it Rossi-XTE} observation, strong X-ray dipping took place (see also Kuulkers
et al. 1998) which can be used to strongly constrain spectral models.
Normally, spectra in several intensity bands would be selected and then
fitted simultaneously. In this case it is difficult to do this
because the spectra were selected from data with 16~s binning
(PCA data in Standard 2 mode) and the ingress to dipping 
was very rapid, on a timescale of seconds (Fig. 3, left). 
To avoid unnecessary mixing of data of different intensities due to
this, which can make spectral fitting difficult and
results erroneous, only two high quality spectra were selected: 
one during deep dipping, and another during persistent emission.

In fitting the dip spectrum, firstly the disc blackbody plus power law
model was used without line terms. All of the emission parameters were held
constant at the values determined for the non-dip spectrum, since dipping 
must be fitted by absorption only. In order to fit the dip spectrum,
it was found necessary for the disc blackbody to have a covering
fraction term implying either partial covering by a blobby absorber,
or steadily increasing, progressive covering by a non-blobby absorber
having angular size somewhat less than the angular size of the disc
blackbody.
In contrast, the power law appeared to be totally covered by a simple 
absorber implying smaller emission region size.  This is the same model as
used by Kuulkers et al. (1998) to fit the same data. However, even the best fit 
for the continuum-only model was poor,
with a reduced $\chi ^2 \sim $2 and so the disc line was added to the
model, for simplicity in the form of two Gaussian lines to represent
the red and blue wings. The model becomes

\vspace{5mm}

\centerline {
$\rm {exp(-\sigma _{pe} {\it N}_{H}^{gal}) \times}$\{[exp(-$\rm 
{\sigma _{pe} {\it N}_{H}^{DBB})\times {\it f}_{cov}+(1-{\it f}_{cov})]\times}$
({\sc discbb}+{\sc 2 Gaussian})+exp(-$\rm {\sigma _{pe} {\it N}_{H}^{PL})
\times}${\sc powerlaw}\}
}

\vspace{5mm}

\noindent where $\sigma _{\rm pe}$ is the photoelectric cross-section 
with cosmic 
abundances (Morrison \& McCammon 1983), $N_{\rm H}^{\rm gal}$ is the galactic 
column density, $N\rm _{H}^{DBB}$ is the disc blackbody
column density,  $f_{\rm cov}$ is the covering fraction
and $N\rm _{H}^{PL}$ is the column density for the Comptonized component.
The best fit model is shown in Fig. 3 (right).
During dipping $N\rm _{H}^{DBB}$ increased to $1.4 \times 10^{24}$ cm$^{-2}$,
the covering fraction  $f_{\rm cov}$ reached 94\% consistent with the
residual intensity seen in dipping (Fig. 3),
and $N\rm _{H}^{PL}$ was much
larger that $N\rm _{H}^{DBB}$ effectively removing all the contribution
of the power law component up to at least 15 keV.  This implies that
the size of the Comptonizing region is smaller that the X-ray emitting
accretion disc.  The intensity of the disc line described for
simplicity by two
Gaussian lines decreased during dipping and could be modelled with the
same covering fraction as the disc blackbody component suggesting that
the line originates in the same part of the disc as the disc blackbody.
The decrease
of the line is highly significant: an F-test showed a significance $>>$ 99.99\%.
This provides further evidence that the disc line is real.

\begin{figure}
\includegraphics[width=90mm, height=87mm]{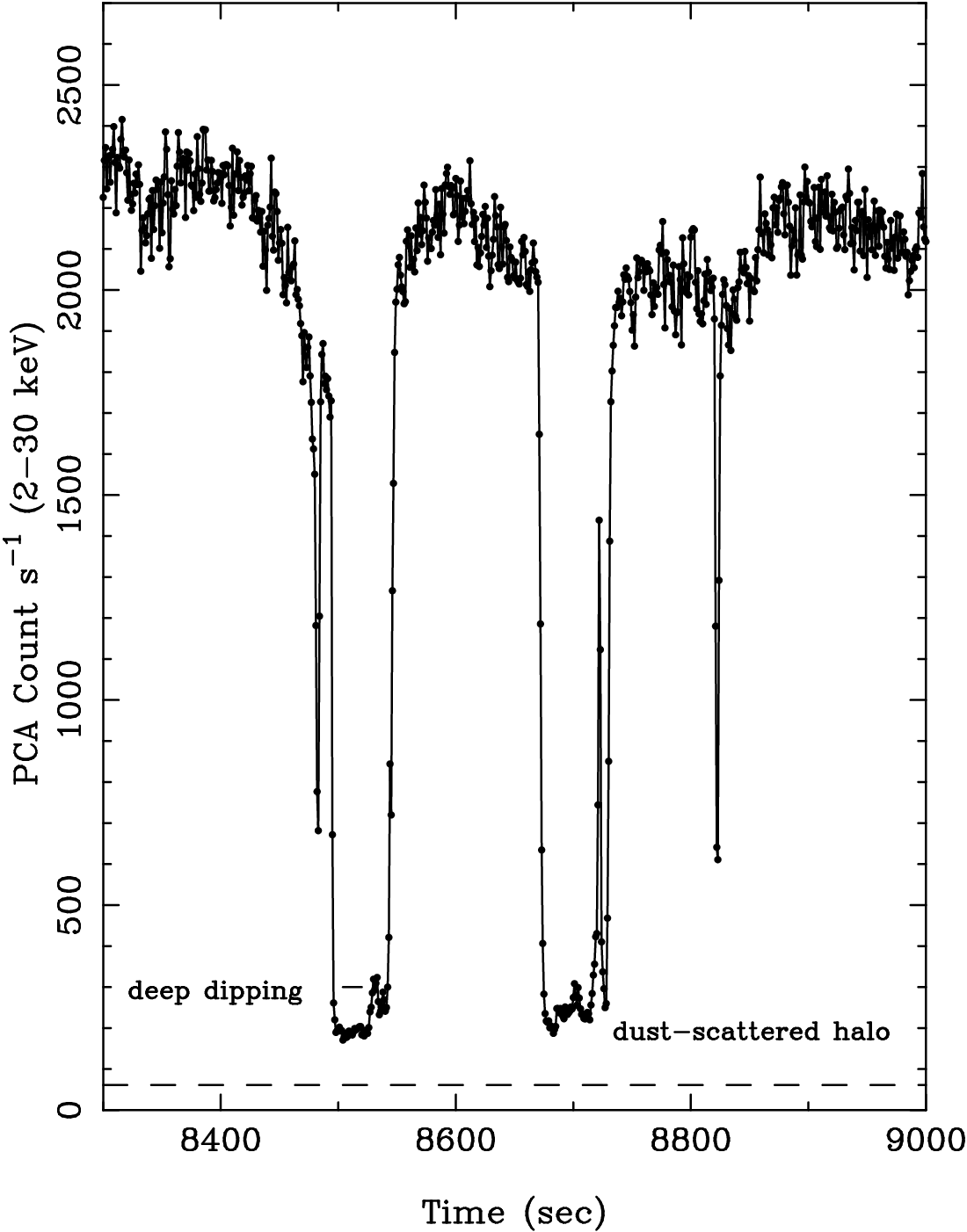}
\includegraphics[width=90mm, height=87mm]{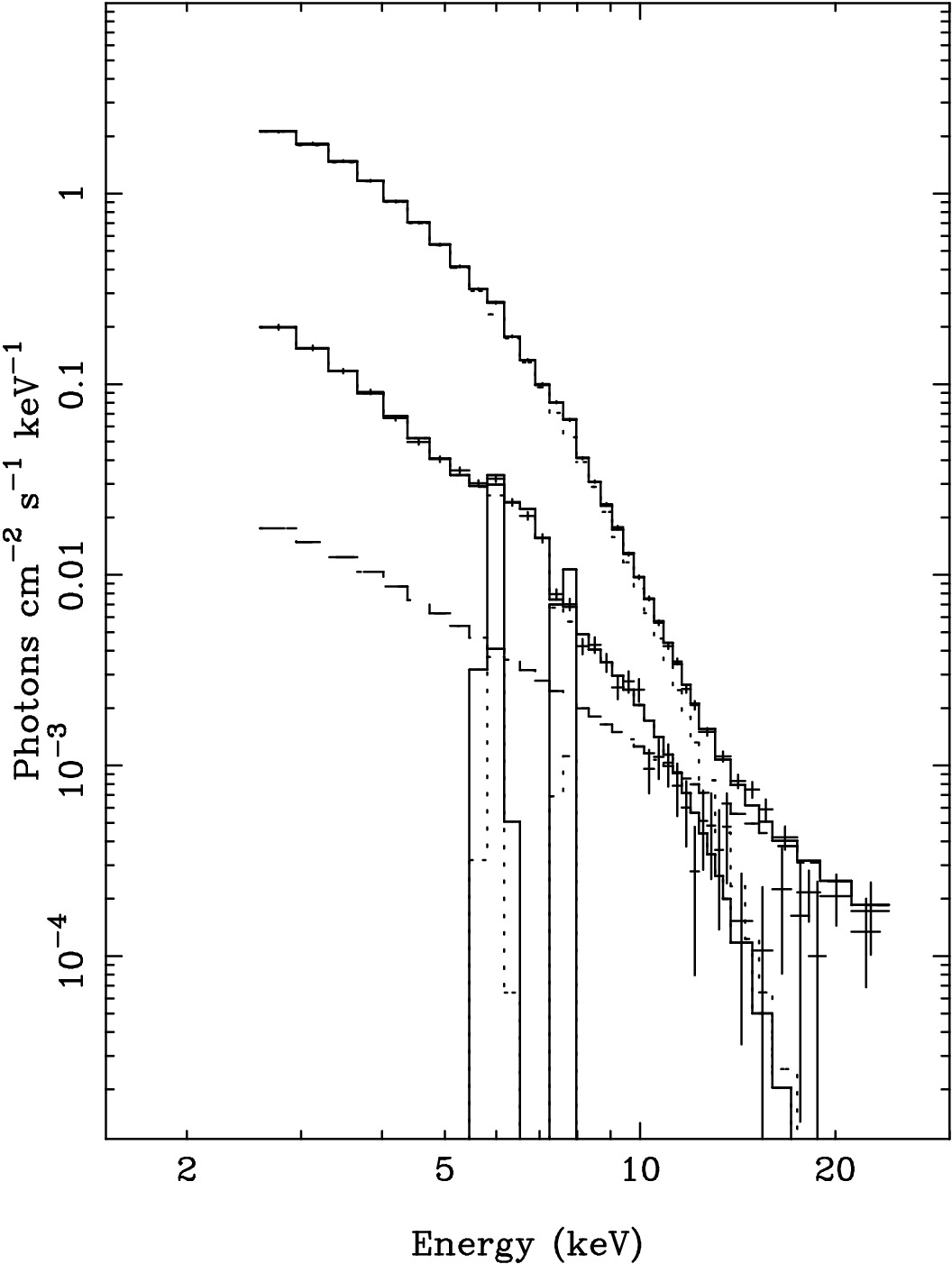}
\vskip -9 mm
\caption{Left: {\it Rossi-XTE} light curve including the dips on February 26, 1997 
with 1-s time
resolution.  The expected count rate (2--30 keV) of the dust-scattered halo
is marked, as well as the section of the data used in the deep dip spectrum.
Right: Unfolded spectra of the persistent and deep dip emission.  The
lines during deep dipping (dotted lines) show a marked decrease.}
\end{figure}%

Although X-ray dipping in \src\ is very deep, it is {\it not} 100\% deep.  
The depth of dipping in the energy band 2.5--25.0 keV is $\sim $91\% 
and is independent of energy if measured in several energy bands.  This rules 
out the possibility that the residual emission comes entirely from
the dust scattered halo.  The halo {\it will} contribute to the residual
emission as expected for a bright source with high galactic column, 
but only at energies below $\sim $5 keV. 
The galactic column
density towards \src\ is $7.0 \pm 0.5 \times 10^{21}$ cm$^{-2}$ based
on $E(B-V) = 1.3 \pm 0.1$ (van der Hooft et al. 1998) and radio measurements
(Dickey and Lockman 1990). From this {\nh}, a dust-scattered
halo contribution of only $\sim $3\% to the count rate in energies 
2.5--25.0 keV can be calculated.
Therefore, the majority of the residual emission during deep dipping must come from 
uncovered disc blackbody emission.  It cannot be Comptonized emission as the total 
contribution of the power law component is 10\% only and spectral
fitting shows that this is totally removed up to at least 
15 keV in deep dipping. Kuulkers et al. (1998) obtain a covering
fraction of 99\% which appears inconsistent with the $\sim $9\% of
intensity remaining in deep dipping (Fig. 3). It is not surprising that
this model produces a low energy excess in dipping; however, we have
shown that this is due to some emission being not covered simply, and
not related to ionized absorber as they suggest.

\section*{SIZE OF THE EMISSION AND ABSORPTION REGIONS}
\begin{figure}
\includegraphics[width=90mm, height=80mm]{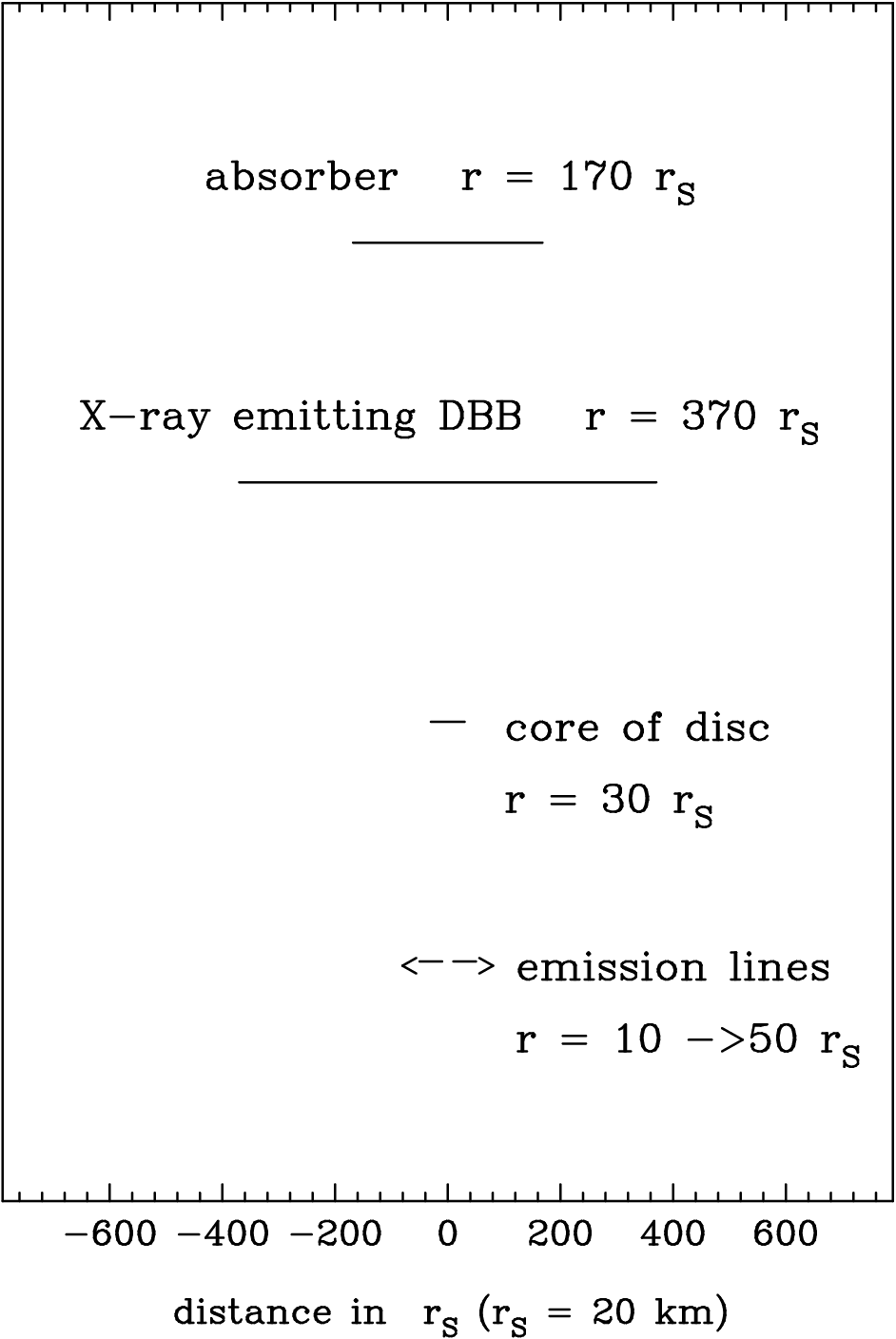}
\includegraphics[width=90mm, height=80mm]{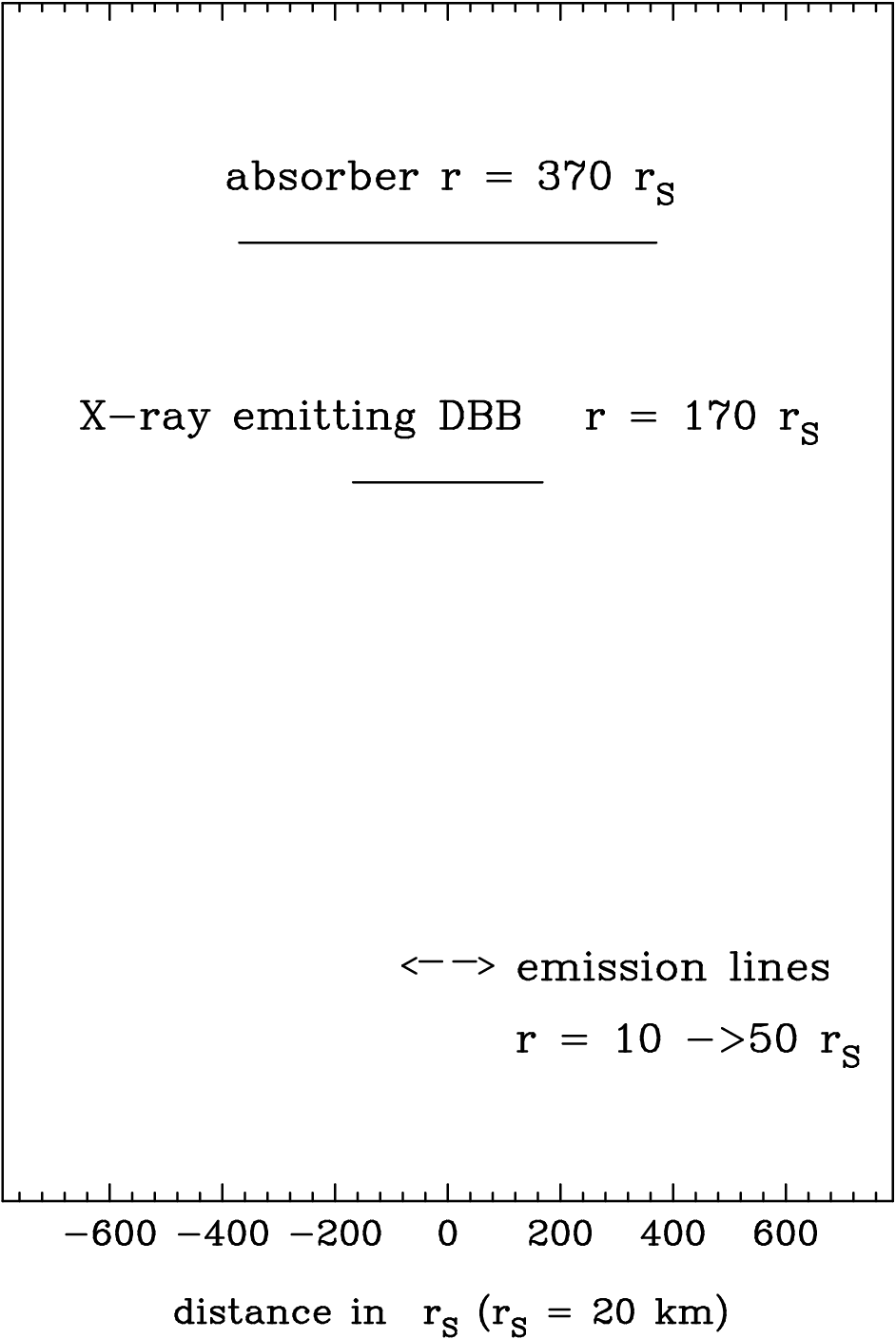}
\vskip -9 mm
\caption{Linear scale of the absorber and the emitter for two cases
discussed.  Sizes are marked in units of Schwarzschild radius.}
\end{figure}%

The sizes of the emission and absorption regions can be obtained by using dip
ingress and egress times, plus the duration of dipping. The technique
differs according to whether the emitter or absorber has larger
angular size, and in the present case that dipping is not 100\% deep 
we cannot tell which is larger.
If the absorber has larger angular size than the most extended emission,
i.e. the disc blackbody, then the ingress time $\Delta t$ will be
determined by the diameter of the emission region and the velocity of
the absorber. The residual emission must be due to the
absorber being blobby. If the absorber has smaller angular size,
$\Delta t$ is determined by the absorber diameter, and the residual due
to incomplete overlap of emitter and absorber. In either case, 
the total duration of dipping including ingress and egress is
proportional to the sum of emitter and absorber diameters.
Both possibilities will be discussed in the following. 
It is assumed that the absorber is located on the outer rim of the disc
and co-rotates with the binary frame.
It should also be noted that changes during dipping take place on
two time-scales: firstly, there is an intensity decrease of $\sim $10--15\%
from the non-dip level of $\sim $2300 Counts~s$^{-1}$
taking $\sim $45~s which can be described as a shoulder of the dipping; 
this is followed by a rapid transition to deep dipping lasting 6--10~s.

Firstly, if the X-ray emitting disc is smaller than the absorber
the total dip ingress time gives the disc blackbody diameter 
$d_{\rm DBB}$ via the equation $2\,\pi \, r_{\rm AD}/{\rm P}$ = 
$d_{\rm DBB}/\Delta t$, where $r_{\rm AD}$ is the accretion disc radius
and $P$ the orbital period. This gives $d_{\rm DBB}$ = 
$\sim 7\times 10^8$ cm. The total duration of dipping of $\sim$170~s 
combined with this value gives an absorber diameter of
$1.5\times 10^9$ cm.

If the emitter is larger, the shoulder and rapid transition to deep
dipping would be interpreted in terms of a core of hot disc blackbody
emitter completely covered in deep dipping, leaving an outer disc
blackbody region uncovered. Thus, the rapid ingress/egress time of 
6--10~s provides the diameter of the core $d_{\rm core}$ = $1.1\times
10^8$ cm. The shoulder ingress time of 45~s gives the absorber diameter
of $7\times 10^8$ cm, and this with the total duration gives 
a total disc blackbody diameter of $\rm {1.5\times 10^9}$ cm.
A simple calculation of the surface brightness of an accretion disc
with temperature of 1 keV confirms that a hot inner core would have a size
of $\sim 10^8$ cm.

The linear scales of the absorber and emitter in the
above cases are presented in Fig. 4. For comparison, the radius
of the region emitting the iron emission line is also shown.
It should be noted that the diameter of the X-ray source obtained here
is substantially larger than that provided by Kuulkers et al. (1998)
of 23 {\rs}. Firstly, they assume that the angular size of absorber
is much greater than that of the source which we show above may not 
be the case, only one possibility. Secondly, they ignored the shoulders 
of dipping lasting $\sim $45~s.
Additionally, they assumed an unrealistically short
ingress/egress time of 3.5~s in comparison with the 6--10~s adopted
here. Consequently, they  obtained a size of the X-ray source 
$\sim $15 times smaller than the value presented here. 
Similarly, the duration of dipping at 55~s was underestimated compared 
with the 170~s used here, leading to radius of $\sim$95 \rs\
compared with our 370 {\rs}.

\section*{CONCLUSIONS}

In the {\it Rossi-XTE} spectra of \src, a highly shifted iron emission line
was detected.  The line originates in the inner part of the
disc which rotates with relativistic velocities $\sim $0.3 c
producing the characteristic shape of a disc line with strong Doppler 
broadening and boosting.
The region where the line is generated is most probably highly
ionized (Fe XXII) although the ionization state is not well
constrained.  The intensity of the emission line decreases
during dipping by the same factor as the emission from
the accretion disc providing further evidence that the line
originates in the disc.

During deep saturated dipping, the Comptonized component is totally
absorbed but 10\% of the emission is still seen which must originate
in the disc.
It can be shown that either the outer and cooler parts of the disc
are still visible during deep dipping or the absorber is blobby.
In both cases, the radius of the disc core which makes the major
contribution to the
disc luminosity is in rough agreement with the radius of the
line emitting region $\sim $30 Schwarzschild radii.  The radius of 
the X-ray emitting region is  between 170--370~\rs\ 
depending on the assumptions made about the absorber. 
The disc emission region is much larger than that
previously reported by Kuulkers et al. (1998), who probably underestimated
the dip ingress/egress times.

\end{document}